\documentstyle[nato,numreferences]{crckapb}

\begin{opening}
\title{GLOBAL CAUSALITY IN SPACE-TIME UNIVERSE}

\author{A.A. Chernitskii}
\institute{Department of Physics, Department of Physical Electronics,\\
St.-Petersburg Electrotechnical University,\\
 Prof. Popov str. 5, St.-Petersburg, 197376, Russia;\\[0.5ex]
A. Friedmann Laboratory for Theoretical Physics\\
St.-Petersburg, Russia.}

\end{opening}

\runningtitle{GLOBAL CAUSALITY IN SPACE-TIME UNIVERSE}

\def\bc{\begin{center}}
\def\ec{\end{center}}
\def\bi{\begin{itemize}}
\def\ei{\end{itemize}}
\def\bin{\begin{enumerate}}
\def\ein{\end{enumerate}}

\begin{document}

\begin{abstract}
The problems connected with a causality of space-time universe
and with the paradox of Einstein, Podolsky, and Rosen
are considered. A main philosophical problem and its possible solutions
are briefly discussed. A concept of unified local field theory
is considered. It is shown that in the framework of such theory
there are nonlocal correlations between space separate events.
These correlations are predicted by quantum mechanics and
they are confirmed by Aspect type experiments for testing of Bell inequality.
The presence of these nonlocal correlations in the framework of
a local field theory is connected with the fact that
its solution is nonlocal in character.
Prospects for possible applications of a unified local field theory
are considered.
\end{abstract}

\setcounter{section}{0}
\setcounter{subsection}{0}

\section{Philosophical Introduction}

In general the problem of causality is closely connected with
basic philosophical problems.

The main philosophical problem concerns a relation between
matter and spirit.
There are two extreme points of view on this problem:
\bin
\item matter is primary and spirit is secondary;
\item spirit is primary and matter is secondary.
\ein
In connection with this topic we should remember the discussion
between Einstein and Tagore \cite{Einstein&Tagor}. In this
discussion Einstein upheld a materialistic view (1.) and
Tagore upheld an idealistic one (2.).

I think that in some sense these extremes can meet or some medium point
of view is possible. But at first I determine some characteristic features
of these extreme views as I understand them:
\bin
\item a materialistic view
\bin
\item there is a common or physical or objective time and space (space-time),
\item there is a unified (physical) law,
\item there are a theoretical predictability for space-time events and
a world order,
\item a person (human) is a part of material world, he has no a free will;
\ein
\item an idealistic view
\bin
\item an individual spirit is free from time and space,
\item any spaces and laws can be formed by consciousness (or consensus),
\item an individual spirit is unpredictable,
\item a person (human) is an incarnation of an individual spirit
which has a free will.
\ein
\ein

A materialistic view  (1.) is confirmed by advances of natural science.
An idealistic view  (2.) can be confirmed by demonstrations of
parapsychology phenomena.

My opinion is that there are space-time with a unified law and
the world order. Being ``normal'' an individual acts in the
framework of the world order
but may be he (his spirit) can also escape (in a certain sense,
sometimes or often) from the material world and space-time.
Thus I think that intermediate states also are possible.

In any case a knowledge of a unified law gives significant
advantages for the civilization, in creation of new technical devices.

Thus the object of my present investigation is a unified law of matter.
The quantum mechanics description can not pretend to this role
because it gives probabilistic predictions in principle.
(In the known expression by Einstein ``God does not play dice''
just a unified law is considered as God.)

In this connection I consider a unified local field theory.

\section{Concept of Unified Local Field Theory of Matter}
\label{sec2}

The whole history of pre-quantum physics naturally led to
the idea of unified field theory for description of matter.
All particles of matter and its
apparent mutual influence must be represented by some solution of
an appropriate field model which must be nonlinear.
Also this model must be local, i.e. it is represented by
some purely differential system of equations.
This is an essence of the ideas which was inspiring
for many scientists in their working.
Let us mention just a few: A.~Einstein, L.~de Broglie, H.~Weil, A.~Eddington,
G.~Mie, E.~Schr\"o\-din\-ger,
M.~Born, L.~Infeld, J.~Plebansky, etc.

But on the other hand the impressive success of quantum mechanics
has eclipsed the idea of unified field theory which was in the air.
The quantum mechanics is essentially a linear theory which is
much more simple for investigation than nonlinear one. But
the quantum mechanics gives the probabilistic predictions only.

Einstein, Podolsky, and Rosen in their famous
article \cite{EPR}
had advanced the arguments for the standpoint that quantum mechanical
description of reality is incomplete.
Bohm and Aharonov \cite{Bohm&Aharonov1957} had proposed an example
(see also my article  \cite{ulftcon})
for demonstration the EPR paradox in which two spin particles scatter
in opposite directions.
According to the quantum mechanical description for this experiment,
a spin states of the individual particles
are indeterminate until a measurement event.
But as soon as we have measured a spin state for one particle
then a spin state for another particle becomes determinate
immediately. This resulting situation is connected with
the conservation law
of full angular momentum for the system of two particles.

This situation looks as though there was an instantaneous
interaction between space separate particles. It
contradicts with the
thesis for locality of interactions.
Thus quantum mechanics predicts nonlocal correlations between the events.
Well known Aspect  experiment \cite{Aspect} for testing of
also well known  Bell inequalities \cite{Bell}
determines that there are the nonlocal correlations.

At first glance the existence of this nonlocal correlations
rejects a possibility for description of matter by
an unified local field theory. However, actually, this is not the fact.
In the following I show that nonlocal correlations
between events must exist in the framework of a unified local
field theory of matter.

\section{Concept of Material World and its Nonlocality}

Actually the concept of unified local field theory for the material world
is similar to the concept of ether, if we understand it
in the broad sense but not a narrow mechanical one.
This concept supposes only two basic properties: continuity and
locality. Mathematically these properties are expressed in the fact that
we consider some purely differential field model or
some system of equations with partial derivatives.
To describe naturally the interactions between material objects,
this system of equations must be nonlinear.
We believe also that there is a model solution
which is determinate in space-time.
Thus, according to this concept, we can
consider some Cauchy problem or the problem with initial condition
for obtaining the world evolution.

Within the framework of such theory a single elementary particle
is represented by some space-localized solution.
Moreover, because, as we know, elementary particles have wave
properties, this solution must have an appropriate wave part.
The wave part is considered here in the sense of time Fourier expansion
for the solution in own coordinate system of the particle,
where this part has the form of a standing wave.

There is a simplest example for such standing wave
even for the customary linear wave equation. These well known solutions
of the wave equation in a spherical coordinate system include
spherical harmonics. For the spherically symmetric case
we have the standing wave
\begin{eqnarray}
&&\frac{\sin (\underline{\omega}\, r)}{\underline{\omega}\, r}\,
\sin (\underline{\omega}\, x^0 )
\end{eqnarray}
which is formed by the sum of divergent and convergent
spherical waves. With the help of Lorentz transformation we can obtain
the appropriate solution in the form of moving
nondeliquescent wave packet.
Then own frequency
$\underline{\omega}$ transforms to wave vector $k_\mu$ such that
\begin{eqnarray}
&&|k_\mu\,k^\mu| {}={} \underline{\omega}^2
\quad.
\end{eqnarray}
Using the linear relation between the wave vector and a vector of momentum
\mbox{$p_\mu= \hbar\,k_\mu$} we obtain
\begin{eqnarray}
&&|p_\mu\,p^\mu| {}={} m^2
\quad.
\end{eqnarray}

A single elementary particle solution of a nonlinear field
model may be called also as {\em solitron}. This term has a similar sense that
``solitary wave'' or ``soliton''. But usually the term ``soliton''
is used in mathematical context for some special solutions.

It is significant,
the concept of unified field theory supposes that
all variety and evolution of the material world
are represented by some space-time field
configuration which is an exact solution of the nonlinear field model.
It is evident that this solution is very very complicated
but it is determinate on space-time by the field model
with initial and boundary conditions.
In the vicinity of a separate elementary particle this world solution
is close to the appropriate single elementary particle solution,
but each elementary particle behaves as the part of the world solution.
Thus the behavior of each elementary particle is connected
with the whole space-time field configuration of the world solution.

For certain conditions it is possible to consider
the world solution part connecting with a separate elementary particle
as the appropriate solitron solution with slowly variable velocity.
(For the case of nonlinear electrodynamics see,
for example, my article \cite{ChernitskiiJHEPb}.)
This level for investigation of the world solution relates to the
classical (not quantum) physics.

It is evident that although the model is local, the world solution
is nonlocal in character because it is determined on a whole space-time
applicable domain.
This means, in particular, that there are undoubtedly
nonlocal correlations between space separate parts
of the common world solution.
This sentence may be explained with the help of the following simplest
example.

\begin{figure}[h]
\begin{center}
\thicklines
\special{em:linewidth 0.4pt}
\unitlength 0.35mm
\linethickness{0.4pt}
\begin{picture}(104.67,90.00)
\put(10,10){\line(0,1){80}}
\put(30,10){\line(0,1){80}}
\put(50,10){\line(0,1){80}}
\put(70,10){\line(0,1){80}}
\put(90,10){\line(0,1){80}}
\put(57,50.00){\vector(1,0){40}}
\put(10.00,50.00){\circle*{2}}
\put(10.00,25.00){\circle*{2}}
\put(40.00,50.00){\circle*{2}}
\put(6,47.00){\makebox(0,0)[rc]{$O$}}
\put(6,24.67){\makebox(0,0)[rc]{$P$}}
\put(40.00,44.50){\makebox(0,0)[ct]{$Q$}}
\put(10.00,5.00){\vector(1,0){90}}
\put(104,5.00){\makebox(0,0)[lc]{$x^1$}}
\put(10,5){\line(0,-1){2}}
\put(40,5){\line(0,-1){2}}
\put(40.00,1.00){\makebox(0,0)[ct]{$q$}}
\put(10.00,1.00){\makebox(0,0)[ct]{$0$}}
\end{picture}
\end{center}
\caption{Customary plane wave.}
\end{figure}
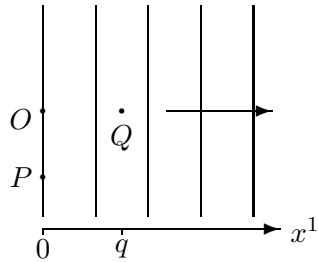

Let us consider a customary plane wave travelling on axis $x^1$
with a fixed wave-length $\lambda$ (see Fig. 1) such that
\begin{eqnarray}
f&=&\sin [(2\pi/\lambda)\,(x^0 {}-{} x^1)]
\quad.
\end{eqnarray}
This wave is the solution of the customary
linear wave equation. At the points $O$, $P$, and $Q$
the field evolution has the forms
\begin{eqnarray}
\nonumber
&&O:\; f=\sin [(2\pi/\lambda)\,x^0]
\quad,
\\
\nonumber
&& P:\; f=\sin [(2\pi/\lambda)\,x^0]
\quad,
\\
\nonumber
&& Q:\; f=\sin [(2\pi/\lambda)\,(x^0 {}-{} q)]
\quad.
\end{eqnarray}

Thus here there are the nonlocal correlations between the field
evolution at the points $O$, $Q$, and $P$.
Totality of such nonlocal correlations is, in fact, the solution
in space-time for the local field model. The possible world solution
(which is extremely more complicated than a plane wave)
is also the continuous set of nonlocal correlations for the field
evolution at the points of three-dimensional space.

Of course, if we make some excitation for field at the point
$O$ then a propagation of this excitation from this point
will have a finite speed. But in the scope of the world
solution we do not be able to make this excitation or
to modify arbitrarily this world solution.
Any excitations of the field at the point $O$
belong to the world solution which is a single whole.
That is, in this case we must consider also all excitations
coming to this point and we will have some standing wave near it.
Thus the world solution is rather a very complicated system of
standing waves than progressing ones.
The initial condition is a common cause of all field excitations
and after a long evolution the different correlations may exist,
even the strange ones.
It can only be said quite positively that the world solution
can be represented by Fourier integral (or series) on orthogonal
space-time harmonics which are essentially nonlocal.
(Here we must remember how
a dominant role is played by orthogonal functions in quantum approach.)

The key to understanding the appearance of momentary distant interaction
in the experiment, which is considered in section \ref{sec2}, is contained
in a concept of chance choice.
Within the framework of the world solution a chance choice is absent,
but both experimenter and experimental apparatus are a part of this
world solution.
That is the orientation of particle spin detectors in the experiment
under consideration is predetermined by the world solution.
We speak about
a chance choice because we do not know the world solution.

As experimentalists, we think that we establish the initial conditions
for the process under investigation but may be this is
too conceitedly and the veritable initial condition was established
earlier. But as theorists, we can already calculate many
correlations between space-time events.

Thus we can suppose that the quantum mechanical description is the level
for investigation of the world solution.
This level take into consideration, in particular,
the global or nonlocal aspects of this solution.

Nonlocality was founded in quantum mechanics from the outset.
In Schr\"o\-din\-ger's picture a free elementary particle
(which have a determinate momentum) is related with
 a plane wave having a constant amplitude on the whole space.
In this case the
quantum mechanical description does not determine a position of
the particle. That is we have the representation of a free elementary
particle by a non space-localized wave that accentuates just nonlocal
aspect of matter.

As we see, there is nonlocality also in the framework of unified
local field theory.
But such theory supposes a solitron model for free elementary particle that
is intuitively more preferable. Furthermore according to this concept
there is a deterministic description of matter.

Thus having a unified mathematical model for matter we
must also take into consideration the whole of the material world evolution
from a start point when the initial condition was determined.
And we can say that there is the global causality in the space-time
universe.

As we see the unified field theory approach can give a strong determinancy
of events in material world. But my opinion is that we
must also take into consideration some things
which are outside from the space-time framework of the physical world.
Suppose here by definition that these things include spirits.
Because a spirit is outside from the material world or a world solution, he
may have an influence on initial and boundary conditions.
Suppose that a spirit can partially modify
the world solution, which is possible with the help of some dynamical
boundary conditions.
But these modifications must be again nonlocal in general.
Thus the world solution is determinate in general
and a possible intervention of spirit must be unexpected
and unusual in character. This spirit intervention also realizes a global
space-time causality in a sense, but this causality is not explained by
physical laws. I think that this is a reasonable way for introduction
a concept of will to the picture of material world stated above.
Thus I believe that in the material world we have a partial
nonlocal determinancy
which however agree with a unified local field theory of matter.

In connection with the fact that I touch on
philosophical problems in this article,
I would like also refer to Schr\"odinger's book
\cite{Schrodinger} including considerations which I accept in general.

\section{Possible Applications of the Unified Local Field Theory}

At present we consider a single atom and even a single electron
as objects of technology. There is a concept of a single electron
transistor \cite{SET} and we can seriously consider prospects
for building an Avogadro-scale computer
acting on \mbox{$\sim 10^{23}$} bits \cite{LloydNature}.
In such computer using
the nuclear magnetic resonance one nuclear spin must store one bit
of information.

Traditional computation can do many useful things and this ability
can become very much stronger with the possible
Avogadro-scale technology. But the traditional computation
needs a determinate controlling. Such controlling is possible if
we have a unified field theory of matter in the sense that was stated
above.

This is one of the possible applications of the approach under review.
But, of course, a realization for the paradigm of
unified field theory will discover
abilities which we do not know at the present time.

In connection with the approach of unified field theory
I propose to consider the nonlinear electrodynamics model
of Born-Infeld type with sin\-gu\-la\-ri\-ties
\cite{ChernitskiiHPA,ChernitskiiJHEPa,ChernitskiiJHEPb,ChernitskiiBICliff}.
In particular, in the framework of this model the two fundamental
long-range interactions (electromagnetism and gravitation)
may be unified (see my articles).


\begin{thebibliography}{99}
\bibitem{Einstein&Tagor} Einstein, A. and Tagore, R. (1931)
The nature of reality, {\em Modern Review (Calcutta)} {\bf XLIX},
42--43.

\bibitem{EPR} Einstein, A., Podolsky, B., and Rosen, N. (1935)
Can quantum-mechanical description of physical reality be
considered complete?,
{\em Phys. Rev.} {\bf 47}, 777--780.
\bibitem{Bohm&Aharonov1957} Bohm, D. and Aharonov, Y. (1957)
Discussion of experimental proof for the paradox of Einstein,
Rosen, and Podolsky, {\em Phys. Rev.} {\bf 108}, 1070--1076.

\bibitem{ulftcon} Chernitskii, A.A. (2002)
Concept of unified local field theory and nonlocality of matter,
 \mbox{quant-ph/0102101}.

\bibitem{Aspect} Aspect, A., Dalibard, J., and Roger G.  (1982)
Experimental test od Bell's inequalities using time-varying analyzers,
{\em Phys. Rev. Lett.} {\bf 49}, 1804--1807.

\bibitem{Bell} Bell, J. S. (1964)
On the Einstein Podolsky Rosen paradox, {\em Physics} {\bf 1}, 195--200.

\bibitem{ChernitskiiJHEPb} Chernitskii, A.A. (1999)
Dyons and interactions in nonlinear (Born-Infeld) electrodynamics,
\mbox{\em J. High Energy Phys.} {\bf 1999}, no. 12, Paper 10, 1--34.

\bibitem{Schrodinger}  Schr\"odinger, E. (1959) {\em Mind and Matter},
Cambridge.

\bibitem{SET} Devoret, M.H. and Schoelkopf, R.J. (2000)
Amplifying quantum signals with the single-electron transistor,
{\em Nature} {\bf 406}, 1039--1046.

\bibitem{LloydNature} Lloyd, S. (2000)
Ultimate physical limits to computation, {\em Nature} {\bf 406},
1047--1054.

\bibitem{ChernitskiiHPA} Chernitskii, A.A. (1998)
Nonlinear electrodynamics with singularities
(modernized Born-Infeld electrodynamics),
{\em Helv. Phys. Acta} {\bf 71}, 274--287.

\bibitem{ChernitskiiJHEPa} Chernitskii, A.A. (1998)
Light beams distortion in nonlinear electrodynamics,
\mbox{\em J. High Energy Phys.} {\bf 1998}, no. 11, Paper 15, 1--5.

\bibitem{ChernitskiiBICliff} Chernitskii, A.A. (2002)
Born-Infeld electrodynamics: Clifford number and spinor
representations, {\em Int. J. Math. \& Math. Sci.} {\bf 31}, 77--84.

\end{thebibliography}
\end{document}